\documentclass[twocolumn,showpacs,preprintnumbers,amsmath,amssymb]{revtex4}

\usepackage{graphicx}
\usepackage{dcolumn}
\usepackage{bm}

\begin{document}

\preprint{APS/123-QED}

\title{ Continuous magnetic phase transition in half-frustrated Ca$_{2}$Os$_{2}$O$_{7}$ }

\author{P. Zheng $^{1}$}
\author{Y. G. Shi $^{1}$}
\author{Q. S. Wu $^{1}$}
\author{G. Xu $^{1}$}
\author{T. Dong $^{1}$}
\author{Z. G. Chen $^{1}$}
\author{R. H. Yuan $^{1}$}
\author{B. Cheng $^{1}$}
\author{K. Yamaura $^{2}$}
\author{J. L. Luo $^{1}$}
\author{N. L. Wang $^{1}$}

\affiliation{$^{1}$Beijing National Laboratory for Condensed Matter
Physics, Institute of Physics, Chinese Academy of Sciences,
Beijing 100080, P. R. China\\
$^{2}$ Superconducting Properties Unit, National Institute for Materials Science,
1-1 Namiki, Tsukuba, 305-0044 Ibaraki, Japan.}%

\date{\today}

\begin{abstract}
We present the specific heat, magnetization, optical spectroscopy
measurements and the first-principle calculations on the Weberite
structure Ca$_{2}$Os$_{2}$O$_{7}$ single crystal/polycrystalline sample.
The Ca$_{2}$Os$_{2}$O$_{7}$ shows a Curie-Weiss nature at
high temperature and goes into a ferrimagnetic insulating
state at $327$ $K$ on cooling. A $\lambda$-like peak is observed at
$327$ $K$ in the specific heat implying a second-order phase
transition. The vanishing electronic specific heat at low
temperature suggests a full energy gap. At high
temperature above the transition, small amount of itinerant carriers
with short life time $\tau$ are observed, which is gapped at $20$
$K$ with a direct gap of $0.24$ $eV$. Our first principle calculations indicate that the anti-ferromagnetic
(AFM) correlation with intermediate Coulomb repulsion U
could effectively split Os(4b) t$_{2g}$ bands and push them away from Fermi level(E$_{F}$). On the other hand, a non-collinear magnetic interaction is needed to push the Os(4c) bands away from E$_{F}$, which could be induced by Os(4c)-Os(4c) frustration. Therefore, AFM correlation, Coulomb repulsion U and non-collinear interaction all play important roles for the insulating ground state in Ca$_{2}$Os$_{2}$O$_{7}$.
\end{abstract}


\maketitle

\section{\label{sec:level2}INTRODUCTION}

Metal-insulator transition (MIT) is one of the important phenomena in
condensed mater physics and has attracted much attention. Many
transition-metal oxides containing partially filled d orbitals, such as NiO, V$_2$O$_3$, have been
identified experiencing such transition
\cite{Bosman,Imada,Qazilbash}. Early in 1949, Mott and
Hubbard proposed that the on-site
Coulomb interaction plays an important role in the narrow
band materials\cite{Mott}. Large on-site Coulomb repulsion energy
could split the band into two. In general, the correlation effect depends on the value
of U$/$W, where W is the band width, and U is the
Coulomb repulsion energy. For 3d transition metal oxides,
the band width W is small and the Coulomb repulsion energy (U) is relatively big,
which could lead to U$/$W$>>1$.
So 3d metal oxides usually belong to strong correlation limit and many of them would be insulating.
For 4d/5d metal oxides\cite{Cox},
their band widths W are larger than that of 3d compounds.
The larger spatial extent of 4d/5d wave functions also means that the on-site Coulomb repulsion is smaller. Therefore 4d/5d materials are generally less strongly correlated than 3d materials, and lead to U/W$\sim$1. Simultaneously, as the atomic mass increasing, the spin-orbital coupling (SOC) becomes more and more important, which also play essential role in 4d/5d materials. With these two effects, the electronic properties of 4d/5d metal oxides become more complex and abundant; many new physical phenomena emerge.

\begin{figure}
\includegraphics[keepaspectratio=true,angle=90,width=0.8\linewidth]{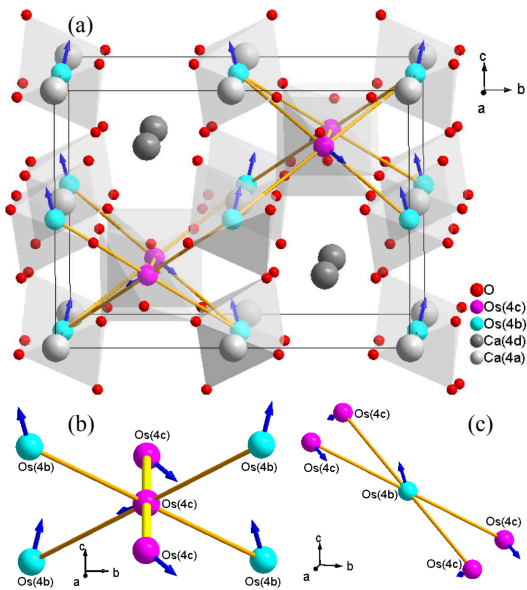}
\caption{\label{fig:epsart}  (a) The crystal structure of Ca$_{2}$Os$_{2}$O$_{7}$, with the 4a-site Ca$^{2+}$ on (0, 0, 0), the 4d-site Ca$^{2+}$ on (1/4, 1/4, 3/4), the 4c-site Os$^{5+}$ on (1/4, 1/4, 1/4) and the 4b-site Os$^{5+}$ on (0, 0, 1/2) in the crystal. (b) Coordination environments of Os(4c) in Ca$_{2}$Os$_{2}$O$_{7}$. (c) Coordination environments of Os(4b) in Ca$_{2}$Os$_{2}$O$_{7}$. A dark blue arrow represents a spin director of Os ion. The orange bonds connect the nearest neighboring Os(4b) and Os(4c) ions. The yellow bonds connect the nearest neighboring Os(4c) and Os(4c) ions. }
\end{figure}

Recently, some partially filled 5d transition metal
oxides, such as Sr$_{2}$IrO$_{4}$\cite{Kim, Ishii},
 Cd$_{2}$Os$_{2}$O$_{7}$\cite{Padilla},
Ca$_{2}$Os$_{2}$O$_{7}$\cite{Reading},
Ca$_{3}$LiOsO$_{6}$\cite{Shi2} et al., were found insulating
at low temperature, which stimulated many experimental and theoretical
studies\cite{Kim, Ishii, Reading, Padilla, Mandrus}. One novel
type of Mott insulating state was observed in the Ir$^{+4}$ based
Sr$_{2}$IrO$_{4}$\cite{Ishii}. The unusual insulating state is
attributed to the cooperative interactions of electron correlation
and large spin-orbit coupling\cite{Kim}. This study is very helpful
to the understanding of the MIT in 5d
transition metal compounds. On the other hand, for the pyrochlore
structure material
 Cd$_{2}$Os$_{2}$O$_{7}$, a continuous
metal-insulator transition was observed by resistance, specific
heat, magnetization, Hall effect, thermal conductivity
 and the optical conductivity measurements\cite{Padilla, Mandrus}. In this
compound, Os 5d t$_{2g}$ bands are half filled because that the
Cd$_{2}$Os$_{2}$O$_{7}$ adopts the octahedral environment of
OsO$_{6}$ with 5d$^{3}$ configuration so that the $t_{2g}$ band is
expected near half filling. As we all know, SOC is weak in S$=3/2$
configuration of the half-filled $t_{2g}$ band for its total orbital
angular momentum L$=$0. One more important issue in
Cd$_{2}$Os$_{2}$O$_{7}$ is that, every Os ion is shared by two Os tetrahedrons,
and frustrated with each other. The large nearest neighbor coordination number
would cause a wider Os band width\cite{Bergman}.  So the mechanism proposed in
Sr$_{2}$IrO$_{4}$ is not suitable for Cd$_{2}$Os$_{2}$O$_{7}$.
Padilla et al attributed this continuous metal-insulator transition in
Cd$_{2}$Os$_{2}$O$_{7}$ to a Slater transition\cite{Padilla}, which
is driven by the AFM correlation. The K$_{4}$CdCl$_{6 }$ type
Ca$_{3}$LiOsO$_{6}$ is another Os$^{5+}$ oxide with the 5d$^{3}$
electronic configuration, in which the t$_{2g}$ orbitals are
half-filled with S$=3/2$. Similarly, the SOC is weak in this case, and it is not
surprising that the calculation with LDA+SOC can not split the Os 5d
t$_{2g}$ bands into two manifolds\cite{Shi2}. However, comparing
with Cd$_{2}$Os$_{2}$O$_{7}$,
 half Os atoms are replaced by Li atoms in Ca$_{3}$LiOsO$_{6}$. The nearest-neighbor Os atoms are rather
distant, and
the magnetic interaction is expected to be rather weak,
regardless of the structural anisotropy.
Therefore, the magnetic frustration is almost quenched.
Band structure calculation shows that the bandwidth of t$_{2g}$ bands
is narrower than for other typical Os
oxides. The smaller bandwidth is a direct
consequence of the local structure around Os atoms\cite{Shi2}.
The narrow t$_{2g}$ bands suggest that
correlation effects are important for this compound,
although the Coulomb interaction is expected to be weak due
to the nature of the 5d orbital.
 Shi et al claim that Ca$_{3}$LiOsO$_{6}$
can be considered as a 5d Mott insulator,
which stabilized within a collinear AFM(anti-parallel) phase\cite{Shi2}.

\begin{figure}
\includegraphics[width=0.85\linewidth]{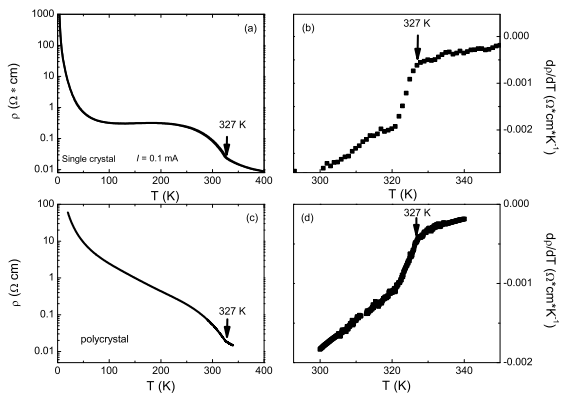}
\caption{\label{fig:epsartff} (a), (b), (c) and (d)  The temperature
dependence of resistivity($\rho$) and d$\rho$$/$dT of
Ca$_{2}$Os$_{2}$O$_{7}$ single crystal and polycrystalline sample
respectively.}
\end{figure}

Ca$_{2}$Os$_{2}$O$_{7}$ is another Os$^{+5}$ oxide with the 5d$^{3}$
electronic configuration but adopts the Weberite
structure\cite{Reading} (see Fig. 1). In this structure, Ca cations
have two different coordinations (4a and 4d) with eight oxygen ions.
Each Os$ ^{5+}$ is octahedrally coordinated by six nearest oxygen
atoms. There are two different Os cation environments. One is
Os(4b) site, which coordinates with four un-frustrated nearest
neighboring Os(4c) cations. The other one is Os(4c) site, which
has six nearest neighboring Os cations. Beside four un-frustrated Os(4b) cations,
it also coordinated by two frustrated Os(4c) cations. So we name Ca$_{2}$Os$_{2}$O$_{7}$ as the half-frustrated system.
The frustration effect
would lead to a non-collinear magnetic interaction\cite{Bergman}. Up to now, only the temperature
dependences of resistivity and crystal structure on polycrystal
were reported. Similar to Cd$_{2}$Os$_{2}$O$_{7}$,
Ca$_{2}$Os$_{2}$O$_{7}$ also has an insulating ground
state\cite{Reading} and shows a metal-insulator-like transition.
However, its magnetic properties, electronic structure and the origin
of the M-I transition are not studied yet. In this work, we present
investigations on Ca$_{2}$Os$_{2}$O$_{7}$ by the resistance, specific
heat, magnetization and optical reflectance measurements. The
resistivity reproduces the behavior reported in ref.\cite{Reading},
which also shows an insulating state at low temperature. Zero
electronic specific heat is observed, implying a full gap at the
low temperature. In the mean time, a spectral weight transferring from
low frequency to high frequency in optical conductivity is observed
with decreasing temperature, conforming the gap opening in the
insulating state. High temperature magnetization study indicates that
the system is stabilized in high spin state with S$=$ $3/2$.
To explain the experimental observations, the first
principles calculations are performed. We find that the AFM correlation with
intermediate Coulomb repulsion U play important roles during Os(4b) bands splitting. But for Os(4c) bands, the non-collinear magnetic interaction is the another necessary factor.

\section{\label{sec:level2}EXPERIMENTS}

High density polycrystal and small single crystal samples were
prepared in a high-pressure apparatus. The stoichiometric mixtures
of CaO , OsO$_{2}$ (Os-$84.0\%$, Alfa Aesar) and KClO$_{4}$ were
placed into a platinum capsule, and sintered in a belt-type
high-pressure apparatus at $1300$ $^{\circ}$C for 1 hour under a
pressure of $3$ GPa. We washed the polycrystal to remove the KCl
flux. The powder samples were pressed again at $6$ GPa to get the
condensed pellet, which were used for measuring the specific heat ,
resistivity, and optical reflectance data. The single crystal grew
under a stable pressure of $3$ GPa during heating at $1500$
$^{\circ}$C for $2$ hours (KCl as flux provided by KClO$_{4}$), then
was slowly cooled down to $1300$ $^{\circ}$C for $3$
hours. Thereafter, the capsule was quenched to ambient pressure and temperature.
The samples were washed by water for several times and we could find
some single crystals. Typically, the crystals have square shape
of surface with edge length about $0.2$ $mm$. The largest single crystal has the edge length about $0.5$ $mm$.
X-ray diffraction measurement at room temperature indicates that
the samples are of single phase with Webrite structure. The crystal structure and main interatomic distances are collected in table. I, which are very close to the previous report\cite{Reading}. The dc
resistivity ($\rho$(T)) and the specific heat(C$_{p}$) measurements
were carried out on a Quantum Design physical property measurement
system (PPMS). The magnetization was measured on a Quantum Design
(SQUID) VSM. The optical reflectance $R(w)$ of
Ca$_{2}$Os$_{2}$O$_{7}$ was measured from $30$ $cm^{-1}$ to $25000$
$cm^{-1}$  at different temperatures on a Fourier transform
spectrometer (Bruker 80v). Standard Kramers-Kronig transformations
were employed to derive the frequency-dependent optical conductivity.
For low frequency extrapolations, we use Hagen-Rubens relation for the measurements
above $300$ $K$ and constant values for reflectance data
below $200$ $K$. In fact, it is found that the different low frequency extrapolations do not affect the conductivity spectra in the measured frequency region. In the high energy
side, the measured reflectance curve is extrapolated constantly to $100 000$ $cm^{-1}$, above which
a well-known function of $\omega$$^{-4}$ is used.

\begin{table}
\caption{\label{tab:table1}The parameters of the crystal structure and main interatomic distances
        at room temperature; group space: $\emph{Imma}$; cell parameters: a $=$ 7.2165${\AA}$, b $=$ 10.1385${\AA}$, c$=$ 7.3844${\AA}$; Ca(4a) on (0, 0, 0), Ca(4d) on (1/4, 1/4, 3/4), Os (4c) on (1/4, 1/4, 1/4), Os (4b) on (0, 0, 1/2), O1 on (0, 1/4, 0.1625), O2 on (0, 0.4038, 0.7279); O3 on (0.2054, 0.3834, 0.4356). }
\begin{ruledtabular}
\begin{tabular}{cc}
\hline

   $ Ca(4a)-O2 $&$ 2.2335 ${\AA}$ $\\
   $ Ca(4d)-O2 $&$ 2.3902 ${\AA}$ $\\
   $  Os(4c)-O1    $&$ 1.9163$ {\AA}$  $\\
  $  Os(4b)-O2    $&$ 1.9451 $ {\AA}$  $\\
$  Os(4b)-Os(4c)    $&$ 3.6176 $ {\AA}$  $\\
$  Os(4c)-Os(4c)    $&$ 3.6082 $ {\AA}$  $\\

\end{tabular}
\end{ruledtabular}
\end{table}

\begin{figure}
\includegraphics[width=0.85\linewidth]{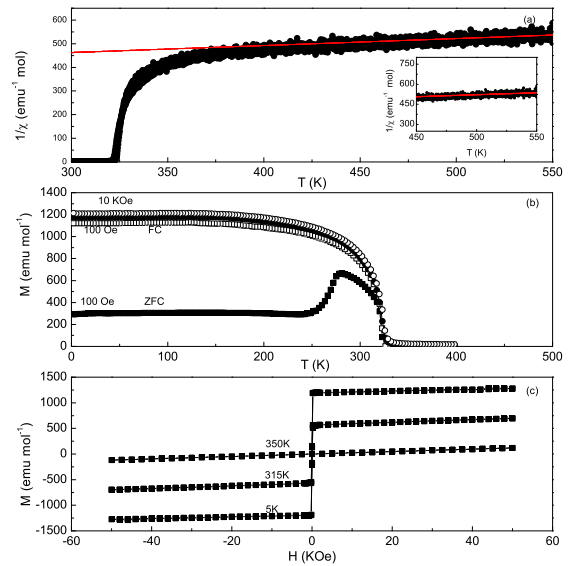}
\caption{\label{fig:epsartfbb}(Color online) (a) The inverse susceptibility as a function of temperature of the polycrystal Ca$_{2}$Os$_{2}$O$_{7}$ between $300$ $K$ and $550$ $K$ in the field of $10$ $KOe$. The inset is $1$$/$$\chi$(T)from $450$ $K$ to $550$ $K$ with a red line representing the Curie-Weiss law fitting. (b) The temperature dependencies of the magnetization of Ca$_{2}$Os$_{2}$O$_{7}$ single crystal in the magnetic fields of $100$ $Oe$(squares) and $10$ $KOe$(circles) after zero field cooling(ZFC)(solid ones) and field cooling(FC)(hollow ones) respectively. (c)The field dependencies of the magnetization of Ca$_{2}$Os$_{2}$O$_{7}$ measured at $350$ $K$, $315$ $K$ and $5$ $K$ respectively.}
\end{figure}

\begin{figure}
\includegraphics[width=0.8\linewidth]{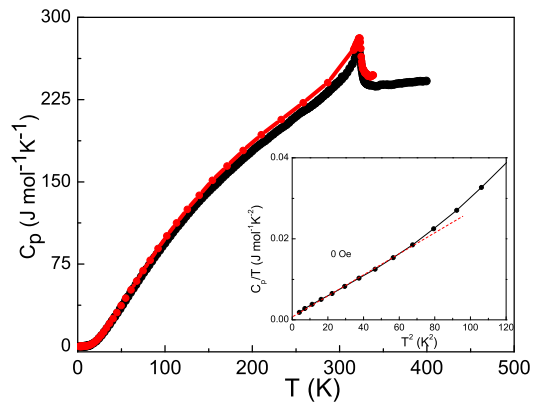}
\caption{\label{fig:epsardgt} The temperature dependencies of
specific heat of ten Ca$_{2}$Os$_{2}$O$_{7}$ single crystals (black $\blacksquare$)
and polycrystal (red $\bullet$). The inset is the enlargement of the low
temperature part of C$_{p}$$/$T vs T$^{2}$.}
\end{figure}

The electronic structures of Ca$_{2}$Os$_{2}$O$_{7}$
are studied by first principle calculations using BSTATE and VASP packages.
Both of them are based on plane-wave method
with pseudopotentail scheme, and give consistent results.
The generalized gradient approximation (GGA)
of PBE-type \cite{Perdew} and its variant GGA+U method
are used for the exchange-correlation energy.
Nonmagnetic (NM), collinear anti-ferromagnetic (AFM)
and non-collinear magnetic states are considered in present study.
All calculations are performed with
the experimental lattice parameters and internal coordinates.
Spin-orbital coupling (SOC) is considered in all calculations.
However, as identified in other  Os$^{5+}$ compounds
such as  Cd$_{2}$Os$_{2}$O$_{7}$\cite{Singh} and
Ca$_{3}$LiOsO$_{6}$\cite{Shi2}, SOC plays very weak roles
on the NM electronic structures and collinear magnetism.

The temperature dependent dc resistivities ($\rho$(T)) for both
single crystal and polycrystalline samples are shown in Fig. 2. They show
similar temperature dependence. A spinodal feature could be found in
$\rho$(T). The spinodals of the two samples appear at the same
temperature(Fig.2 (a) and (c)). The resistivity at $300$ $K$ is ten
times higher than the value for Cd$_{2}$Os$_{2}$O$_{7}$\cite{Mandrus} but one
percent of the value for Ca$_{3}$LiOsO$_{6}$\cite{Shi2}. We
calculate their $d\rho/d$T(Fig.2 (b) and (d)) to determine the
phase transition temperature. The sharp drop of $d\rho/d$T begins at
$327$ $K$ for both samples.  $d\rho/d$T of
the single crystal changes more pronouncedly than that of polycrystal. Both the XRD result and
the $\rho$ $(T)$ imply the high quality of our samples and that the
polycrystalline sample is very close to the single crystal.

The magnetic properties of Ca$_{2}$Os$_{2}$O$_{7}$ under magnetic fields are shown in Fig.
3 (a) (b) and (c). Figure 3 (a) is the
inverse susceptibility ( $1/\chi$(T) ) from $300$ $K$ to $550$ $K$. An
abrupt change occurs at the same critical temperature with the
resistivity anomaly. The data between $320$ $K$ and $400$ $K$
deviate from the linear-T dependence. When temperature is higher
than $400$ $K$, $1$$/$$\chi$ shows a good linear relation with T,
suggesting a paramagnetic Curie-Weiss behavior,
$\chi$ $=$ $N\mu_{0}\mu_{J}^{2}$$/3k_{B}$(T $-$ $\theta$),
from which we can estimate the system's intrinsic magnetic moment, in terms of
C$=$$N\mu_{0}\mu_{J}^{2}/3k_{B}$. Here $\theta$ is the Weiss
temperature; N is the Avogadro¡¯s constant; $k_{B}$ is the Boltzmann constant;
$\mu_{J}$ $=$ g$\sqrt{S(S+1)}\mu_{B}$ (Lande factor g$=$ 2) is the intrinsic magnetic moment. By fitting
the data from $450$ $K$ to $550$ $K$ ( shown in the inset of Fig.
3(a)), the intrinsic magnetic moment per Os ion is determined to be $2.59$$\mu_{B}$. This value is about $67$$\%$ of
the expected moment($3.87$$\mu$$_{B}/$Os) for S$=$ $3/2$ configuration,
indicating that the spins of Os 5d electrons are in the S$=$$3/2$
high spin state. $\theta$ is $-1263$ $K$, suggesting a strong
antiferromagnetic interaction.

Figure 3 (b) shows the temperature dependences of the magnetization
M(T) measured with $100$ $Oe$ and $10$ $kOe$ field respectively.
Each measurement was performed under zero field cooling(ZFC) and field
cooling(FC) respectively. Steep change of M(T) at $327$ $K$ could be
observed in all the curves, suggesting the establishment of a long
range magnetic ordering at T $=$ $327$ $K$. The ZFC data measured
under 10kOe is the same with the FC data. Since high magnetic field
may make the spins polarized, the intrinsic magnetic properties of
the system may not be completely revealed. Therefore, a small field
$100$ $Oe$ ZFC is used, which may reflect more intrinsic information
of the system. Looking at the ZFC M(T) curve under $100$
$Oe$, we can see that the magnetization keeps a small constant value
until $250$ $K$, and increases to the maximum at about $280$
$K$. It drops abruptly at about $327$ $K$ and decreases weakly with
T increasing further. Such a ZFC M(T) is the typical behavior of the ferrimagnetic
ordering or the canted AFM ordering with small net moments.
From the data measured at $5$ $K$ shown in Fig. 3(c), we get the
magnetization to be $1200$ $emu$ $mol^{-1}$ at the largest field $50$ $kOe$,
equaling to $0.107$$\mu$$_{B}$ per Os.
The net moment is only $3.6\%$ of the expected moment for S$=$
$3/2$.

\begin{figure}
\includegraphics[width=0.8\linewidth]{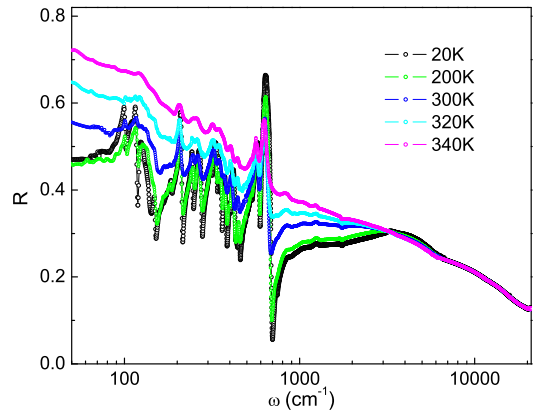}
\caption{\label{fig:epsartgsjjk}(Color online) The frequency dependence
of reflectivity $R(w)$ at $340$ $K$, $320$ $K$, $300$ $K$, $200$ $K$ and $20$ $K$ in the frequency region from
$50$ $cm^{-1}$ to $22,000$ $cm^{-1}$, respectively.}
\end{figure}

In Fig. 4, we plot C$_{p}$ vs T for single crystal( represented by
black squares) and polycrystal (represented by the red circles). The
experimental curves of polycrystal and single crystal have the same
shape and show phase transition at the same temperature. For both
samples, the specific-heat anomaly starts at about $327$ $K$ and
gets to a maximum at $323$ $K$. It corresponds to the transition
observed in the resistance and magnetic measurements. The shape of
the specific-heat anomaly is $\lambda$-like, indicating that the
transition is a second-order phase transition. This behavior is
consistent with that no crystal structure phase transition was observed in
the variable temperature crystal structure analysis in ref. \cite{Reading}.
The inset of Fig. 4 shows
the low-temperature part of C$_{p}$$/$T vs T$^{2}$ (solid line)
and the fitting result(dashed line) using formula C$_{p}$(T)$/$T $=$ $\beta$T$^{2}$ $+$
$\gamma$.
One can see that the function could well fit the data at T$<$ $7$ $K$. The T$^{3}$ term
in C$_{p}$(T) could be attributed to either phonon or/and magnon
contributions. The linear term is the contribution from the
electrons. We found that the low-temperature C$_{p}$$/$T approach
a small value ($<$ $1.4$ mJ mol$^{-1}$ K$^{-2}$ ) in the limit of
T$\rightarrow$0, indicating the vanishing of the density of states. It
implies that the system is fully gapped.

\begin{figure}
\includegraphics[width=0.8\linewidth]{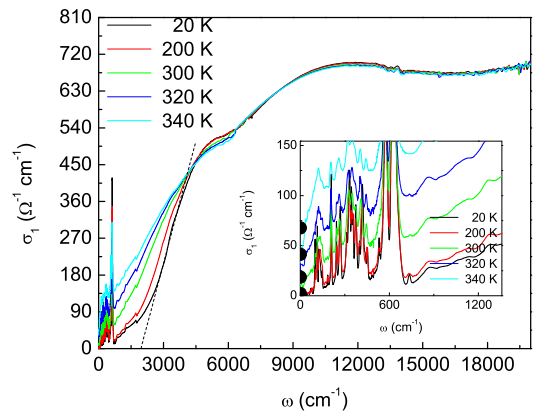}
\caption{\label{fig:epsarthshe}(Color online) Frequency dependence of
the optical conductivity $\sigma_{1}(\omega)$ at different
temperatures. The dashed line shows the steepest dropping of the conductivity and crossing the $\omega$-axis which is defined as the direct gap $2\Delta$. The inset is the enlarged scale of $\sigma_{1}(\omega)$ in which the black points are the dc resistivities at the relative temperatures .}
\end{figure}

To understand the phase transition, we
have to know the electronic structure of Ca$_{2}$Os$_{2}$O$_{7}$.
Optical spectroscopy is a powerful tool to probe the electronic
structure and charge dynamics. We measured the
reflectivity spectra from $30$ $cm^{-1}$ to $22000$
$cm^{-1}$ at various temperatures and plotted them in
Fig. 5. Figure $6$ is a collection of the real part of the optical
conductivity $\sigma_{1}(\omega)$ between $20$ $K$ and $340$ $K$.
The black points in the inset are the dc conductivity from Fig.2 (c), which match with the
lowest frequency optical conductivity very well. We can see that there are a lot
of phonon responses in the optical conductivity spectra. Even at $340$ $K$, the conductivity decreases with decreasing $\omega$. Although a Drude-like response is not visible, the value of $\sigma_{1}(\omega)$ at the lowest frequency does not vanish. The data indicate that the compound may has a few itinerant carriers at high temperature, which contribute to a small Drude component. A quantitative analysis will be presented below. With decreasing T, the phonon peaks become more pronounced.
We found that some modes, for example $577$ $cm^{-1}$ and $641.5$
$cm^{-1}$, weakly harden when $T$ $<$ $320$ $K$, indicating some bonds weakly shrinking in the crystal structure\cite{Reading}.

The conductivity spectra display two inter-band transition peaks at about
$5000$ $cm^{-1}$(about $0.6$ $eV$) and $11000$ $cm^{-1}$ (about
$1.3$ $eV$). A full gap could be
clearly observed below $200$ $K$. This is also consistent with the
zero electron specific heat in Fig. 4. With the temperature decreasing, the spectral weight transfers from low frequency to high frequency and a direct gap opens gradually. To determine the energy gap at $20$ $K$, we draw a dashed line to fit the steepest part of
$\sigma_{1}(\omega)$ in Fig. 6.
Extending the line to the $\omega$-axis, the crossing appears at $2\Delta$ $\approx$ $2000$ $cm^{-1}$, which corresponds to the direct gap. Using the same way, we determine the gap at different temperatures and collect them in Fig .7, showing how the direct gap develops with the temperature varying. A relatively large gap opens at
the same temperature as that shown in the magnetization and specific heat measurements. Consistent with a strongly correlated material, the gap opens at a temperature much lower than the size k$_{B}$T$_{AFM}$ $<<$ 2$\Delta$.

\begin{figure}
\includegraphics[width=0.8\linewidth]{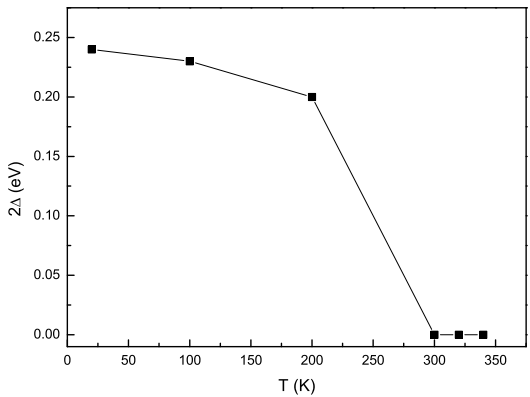}
\caption{\label{fig:epsartfdhr}(Color online) The development of the direct gap with temperature. }
\end{figure}

We noticed that at T$>$ 300 $K$, the optical conductivity spectra do not simply follow the Lorentz lineshape in the low frequency region. As described above, an abrupt increasing reflectivity with the frequency decreasing could be clearly observed in the far infrared region. When the gap completely opens, the reflectivity is almost frequency independent as shown in Fig. 5. We used the Drude -Lorentz model to fit the reflectivity at $340$ $K$ and $20$ $K$, respectively:

\begin{eqnarray}
  \varepsilon(\omega)=\varepsilon_{\infty}-\frac{\omega_{p}^{2}}{\omega^{2}+i\gamma\omega}+\sum_{j}^{N}\frac{S_{j}^{2}}{\omega_{0j}^{2}-\omega^{2}-i\gamma_{j}\omega}.
\end{eqnarray}

where $\varepsilon$$_{\infty}$ is the high-frequency dielectric constant; the
middle and last terms are the Drude and Lorentz components,
respectively. The fitting results are shown in Fig.8. A Drude term and two Lorentz terms were used to fit the reflectivity at $340$ $K$ and $20$ $K$ respectively, which match the raw data very well. The parameters are listed in Table.II. This analysis leads to $\omega_{p}$$=$4383 cm$^{-1}$ and
the scattering rate 1$/$$\tau$$=$2572 cm$^{-1}$ for the Drude component at $340$ $K$. At $20$ $K$, the Drude component becomes vanishing; thus, two Lorentz terms could roughly reproduce the reflectivity as shown in the down-part of Fig.8. The  fitting results confirm that there should be some short-life time itinerant carriers in the system at $340$ $K$, which is completely gaped at $20$ $K$.

\begin{table}
\caption{\label{tab:table2}The parameters of function(1)
        at 340 $K$ and 20 $K$}
\begin{ruledtabular}
\begin{tabular}{ccccc}
 $T(k)$&$ function $&$\omega_{0}(cm^{-1})$&$\omega_{p}(cm^{-1})$&$\gamma (cm^{-1})$\\
\hline
  $ 340K $&$ Drude $&$ 0  $&$ 4400 $&$ 2600 $\\
  $      $&$ Lorentz1 $&$ 4500 $&$ 12600 $&$ 9300 $\\
  $      $&$ Lorentz2 $&$ 12300$&$ 22000  $&$  14900  $\\
  $ 20K $&$ Drude $&$ 0  $&$ 0 $&$ 0 $\\
  $      $&$ Lorentz1 $&$ 4500 $&$ 13600 $&$ 9300 $\\
  $      $&$ Lorentz2 $&$ 12300 $&$ 22000 $&$ 14900 $\\

\end{tabular}
\end{ruledtabular}
\end{table}

\begin{figure}
\includegraphics[width=0.8\linewidth]{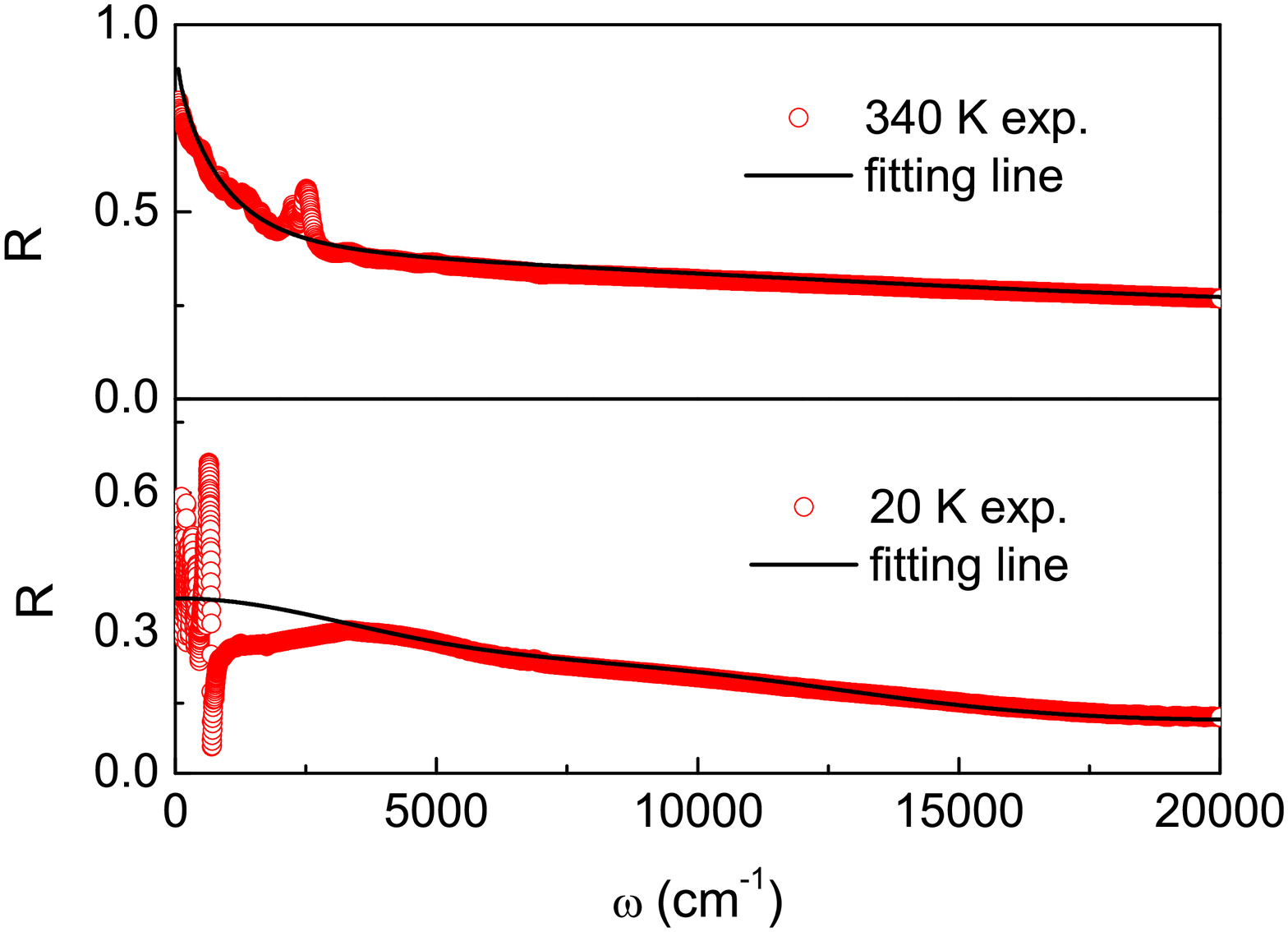}
\caption{\label{fig:epsartgsjjk}(Color online) Fitting the frequency dependence
of reflectivity $R(w)$ at $340$ $K$ and $20$ $K$ respectively. The circles represent the raw data and the solid line is the fitting line using the Drude-Lorentz model.}
\end{figure}

\section{\label{sec:level2}CALCULATION AND DISCUSSIONS}

\begin{figure}
\includegraphics[width=0.8\linewidth]{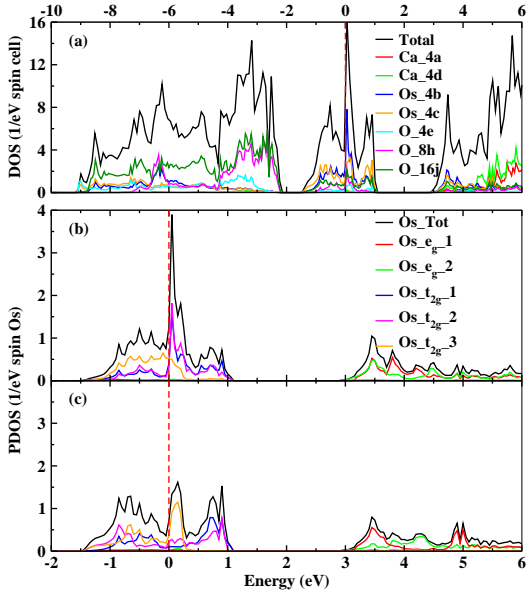}
\caption{\label{fig:DOS-NM}(Color online) The calculated electronic properties
of NM states without SOC.  (a) is the DOS and PDOS in a primitive cell.
(b) and (c) are the PDOS on 5d orbitals of Os(4b) and Os(4c).
Fermi level E$_F$ is defined at 0, denoting as the vertical red dash line. }
\end{figure}

The electronic structures of Ca$_{2}$Os$_{2}$O$_{7}$
are calculated using density functional method.
Because SOC has little influence on the Os$^{5+}$ compounds,
we only show the total and projected densities of states
(DOS and PDOS) for the nonmagnetic calculation without SOC in Fig. 9.
From Fig. 9 (a) we can see that the t$_{2g}$ and e$_{g}$ manifolds are clearly
separated from each other due to the octahedral crystal field effect,
and from the O 2p bands by clean gaps.
As a result, the t$_{2g}$ manifolds mostly occupy from -1.5 $eV$ to 1.1 $eV$
for both Os 4b and 4c sites as shown in  Fig.9 (b) and (c),
well separated from the unoccupied e$_{g}$ manifolds
which are strongly hybridized with O 2p orbitals
and localized between 3 and 5 $eV$.
We can see that the t$_{2g}$ manifolds are half filled
and contain 12 bands, confirming that
there are 2 Os(4b) and 2 Os(4c) ions per unit cell,
and each Os ion's valence is exactly +5.
The t$_{2g}$ band widths for both Os(4b) and Os(4c) are approximately 2.6 $eV$,
very similar to 2.85 $eV$ of Cd$_{2}$Os$_{2}$O$_{7}$ \cite{Singh},
but much wider than Ca$_{3}$LiOsO$_{6}$'s 1.2 $eV$ \cite{Shi2}.
This means that: 1) the 4b and 4c sites have the similar local conditions
around Os ions from the nonmagnetic view;
2) the on-site Coulomb interaction itself could not
be responsible for Ca$_{2}$Os$_{2}$O$_{7}$'s insulating ground state;
3) the distance between the nearest Os neighbors $d_{Os-Os}$
plays an important role for their t$_{2g}$ band width in the Os$^{5+}$ compounds.
For example, the average $d_{Os-Os}$ of Ca$_{2}$Os$_{2}$O$_{7}$
is about 3.605 $\AA$, very close to Cd$_{2}$Os$_{2}$O$_{7}$'s 3.592 $\AA$,
but much smaller than that in Ca$_{3}$LiOsO$_{6}$ (about 5.648 $\AA$).
So that the former two have the similar t$_{2g}$ band widths,
while the latter one is much narrower than them.

\begin{figure}
\includegraphics[keepaspectratio=true,angle=90,width=0.9\linewidth]{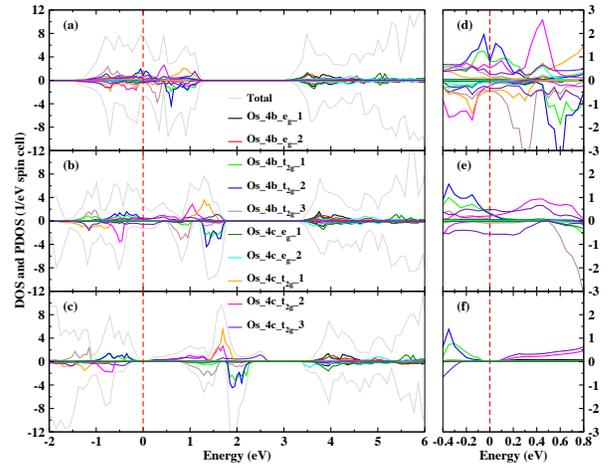}
\caption{\label{fig:DOS-AFM}(Color online) The calculated electronic properties
of the collinear AFM states without SOC by GGA (+ U) method.
(a), (b) and (c) are the results calculated by U $=$ 0 (GGA), 2 and 4 $eV$ respectively.
The positive and negative show partial DOS for up and down spins, respectively.
Vertical red dash denotes the Fermi level E$_F$. (d), (e) and (f) are the enlargements of the Fermi level region of (a), (b) and (c) respectively.   }
\end{figure}

It should be noted that the high value of DOS at E$_{F}$
(N(E$_{F}$) = 13.203 $eV$$^{-1}$ per cell ) obtained in
Ca$_{2}$Os$_{2}$O$_{7}$ would lead to strong magnetic instability
in this material. In order to investigate the possible magnetic order
and the insulating ground state, we have performed GGA+U calculation for a collinear AFM state, where the moments on 4b and 4c sites are assumed to be opposite.
The results are shown in Fig. 10, from which we can see that
the N(E$_{F}$) of GGA is reduced to 7.646 $eV$$^{-1}$ per cell when collinear AFM is considered,
and the total energy is about 155 meV/cell lower than NM state.
With U increasing, the up and down spin orbitals on one site will further split,
leading to the reduction of N(E$_{F}$). An insulating state is reached at U = 4 $eV$.
Such a large U is obviously unreasonable for a 5d metal oxide.
However, if we carefully examine the process of gap opening,
some useful clues become clear. One important character of Fig.10 is that
the collinear AFM correlation can obviously narrow the t$_{2g}$ bands on 4b-sites,
while has weak effect on 4c-sites. Taking the majority channel
(up spin for 4b and down spin for 4c) calculated by GGA as an example,
the band width of 4b becomes 2 $eV$ ranging from -1.5 to 0.5 $eV$,
but the band width of 4c remains 2.5 $eV$ localized between -1.5 and 1.0 $eV$.
Thus, when the Coulomb repulsion U increases to 2 $eV$, a full gap opens
between up and down spin channels of 4b-sites as shown in Fig.10 (b),
while the 4c-sites remains metallic until the unreasonable U = 4 $eV$ is used.
These different behaviors of the electronic structures between 4b and 4c
remind us their different conditions of the nearest neighbor Os ions.
In the view of the nearest Os-Os spin interaction,
one 4b-site spin is coordinated by four opposite spins on 4c-site.
However, for the 4c-site, besides the AFM interaction with four opposite spins on 4b-site,
it is frustrated with other two 4c-site spins,
$i.e.$ the spins between 4c-sites have ferromagnetic interaction with each other.
In this circumstance, we think that the electronic properties of 4b-site Os ions
in Ca$_{2}$Os$_{2}$O$_{7}$ is very similar to NaOsO3 \cite{Shi3},
where the gap opening is duo to the dual effect of AFM correlation and Coulomb repulsion.
On the other hand, the physics on 4c-site is  more or less like
what happened in Cd$_{2}$Os$_{2}$O$_{7}$,
where magnetic frustration play an important role.
Considering our resistance and optical measurements that
Ca$_{2}$Os$_{2}$O$_{7}$ undergoes a wide temperature region
to transform into the real insulating state, we identify that
the electronic structures on 4c-site experience a Slater transition.

\begin{figure} [tp]
\includegraphics[width=0.9\linewidth]{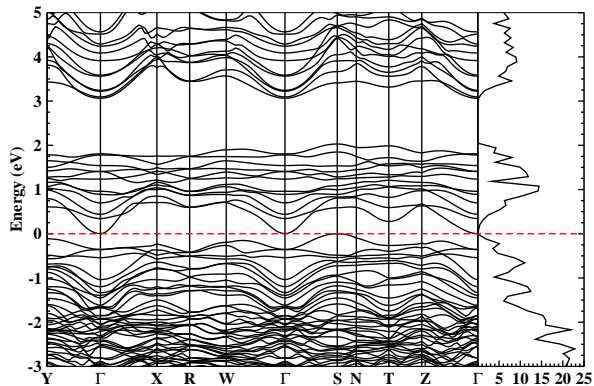}
\caption{\label{fig:epnoncol}(Color online) The calculated electronic properties
of the non-collinear magnetic states by GGA + U method with SOC and U = 2.5 $eV$.
The left panel is the band structure, where the high symmetry points
are defined as Y(0.5,-0.5,0.5), $\Gamma$(0,0,0), X(-0.377,0.377,0.377),
R(0,0,0.5), W(0.25,0.25,0.25), S(0,0.5,0.5), N(0.255,0.378,-0.005),
T(0.5,0,0) and Z(0.383,0.383,-0.383),
and the right panel is the corresponding DOS with unit 1/ ($eV$ cell)
Red dash denotes the Fermi level E$_F$}
\end{figure}

   In order to find the real magnetic states,
magnetic frustration must be correctly dealt with.
Thus we perform GGA $+$ U method with SOC to calculate
the non-collinear magnetic states of Ca$_{2}$Os$_{2}$O$_{7}$,
and obtain an insulating solution with a plausible Hubbard U (2.5 $eV$).
The results are shown in Fig. 11, from which we find that
a direct gap about $0.35$ $eV$ is opened, very close to
the data (0.24 $eV$) deduced by optical measurements.
Moreover, based on the non-collinear magnetic DOS,
the inter-band transition at $5000$ $cm^{-1}$ (about $0.6$ $eV$)
and $12000$ $cm^{-1}$ (about $1.5$ $eV$) in $\sigma_{1}$$(\omega)$
correspond to the hopping from -0.2 $eV$ to 0.6 $eV$ and -0.6 $eV$ to 1.0 $eV$ very well.
Notes that the Os 5d t$_{2g}$ orbitals are hybridized with O 2p orbitals,
and the moments (spin) are non-collinear now, so that the hopping could happen.
The calculated moment direction for each Os is shown as arrow in Fig.1.
We think that the measured net magnetic moments at low temperature
are derived from the non-collinear arrangement on Os.

\section{\label{sec:level2}CONCLUSION}

   We report our specific heat, magnetization and the optical
reflectance investigations on Ca$_{2}$Os$_{2}$O$_{7}$. The phase transition
is revealed by all the experiments. An energy gap is observed
developing with decreasing temperature in the optical measurements. Along with the gap opening, a net ZFC
magnetic moment appears and the spectral
weight in low frequency region is transferred to higher frequency region.
We performed the first principle calculations and found that the anti-ferromagnetic
(AFM) correlation with intermediate Coulomb repulsion U
could effectively split Os(4b) t$_{2g}$ bands and push them away from Fermi level (E$_{F}$), while a non-collinear magnetic interaction is needed to push the Os(4c) bands away from E$_{F}$, which could be induced by Os(4c)-Os(4c) frustration.

\begin{acknowledgments}
This work is supported by National Science Foundation of China( No. 10834013, No. 10874213),
the 973 project of the Ministry of Science and Technology of
China, and Grant-in-Aid
for Scientific Research of No. 22246083 from JSPS .
\end{acknowledgments}

\end{document}